\documentstyle[aps,multicol,epsfig,epsf,psfrag]{revtex}

\draft
\newcommand{\beq}{\begin{equation}}
\newcommand{\eeq}{\end{equation}}
\newcommand{\bdis}{\begin{displaymath}}
\newcommand{\edis}{\end{displaymath}}
\newcommand{\bea}{\begin{eqnarray}}
\newcommand{\eea}{\end{eqnarray}}
\newcommand{\barr}{\begin{array}}
\newcommand{\earr}{\end{array}}

\begin{document}

\title{Bursts in a fiber bundle model with continuous damage}

\author{Raul Cruz Hidalgo$^{1}$, Ferenc Kun$^{1,2}$\footnote{Electronic 
address:feri@dtp.atomki.hu}, and Hans. J. Herrmann
  $^{1}$} 

\address{$^1$ICA 1, University of Stuttgart, 
Pfaffenwaldring 27, 70569 Stuttgart, Germany\\
$^2$Department of Theoretical Physics, University of Debrecen, \\ 
P.O.Box: 5, H-4010 Debrecen, Hungary}

\date{\today}
\maketitle
\begin{abstract}
We study the constitutive behaviour, the damage process, and the
properties of bursts in the continuous damage fiber  
bundle model introduced recently. Depending on its two parameters, the
model provides various types of 
constitutive behaviours including also macroscopic plasticity. 
Analytic results are obtained to characterize the damage process along
the plastic plateau under strain 
controlled loading, furthermore, for stress controlled experiments we
develop a simulation technique and explore numerically the
distribution of bursts of fiber breaks assuming infinite range of
interaction. Simulations revealed that under certain conditions power
law distribution of bursts arises with an exponent significantly
different from the mean field exponent $5/2$. A phase diagram of the
model characterizing the possible burst
distributions is constructed.
\end{abstract}

\pacs{PACS number(s): 46.50.+a, 62.20.Fe, 62.20.Mk}

\begin{multicols}{2}
\narrowtext

\section{introduction}

Recently, the breakdown of disordered materials under externally
imposed stresses has attracted much attention
and by now several aspects of the breakdown process are well
understood \cite{hans,chakrab}. The possibilities of pure analytical
approaches for 
breakdown phenomena are rather limited, hence, computer simulation is
an indispensable tool in this field. Models, computer simulations are
based on, can be classified as lattice models and fiber bundle models.   
In lattice models the elastic medium
is represented by a spring (beam) network, and disorder is captured
either by random dilution or by assigning random failure thresholds to
the bonds \cite{hans}. The failure rule usually applied in lattice
models is discontinuous and irreversible: when the local load 
exceeds the failure threshold of a bond, 
the bond is removed from the calculations ({\em i.e.} its
elastic modulus is set to zero). 

A very important class of models of material failure 
are the fiber bundle models (FBM) \cite{daniels,coleman,krajcinovic,sornette1,sornette2,bernardes,moukarzel,roux,yamir1,kun1,yamir2,phoenix1,phoenix2,phoenix3,phoenix4,kun2,beyerlein,curtin1,curtin2,curtin3,hild,leath,delaplace,curtin4,curtin5,curtin6,hansen1,hansen2,hansen3,newman1,newman2,duxbury}, 
which have been extensively
studied during the past years. These models consists of a set of
parallel fibers having statistically distributed strength. 
The sample is loaded parallel to the fibers direction, and the fibers 
fail if the load on them exceeds their threshold value. In stress 
controlled experiments, after each fiber failure the load carried by the 
broken fiber is redistributed among the intact ones. The behaviour of
a fiber bundle under external loading strongly depends on the range of
interaction, {\it i.e.} on the range of load sharing among 
fibers. Exact analytic results on FBM have been achieved in the
framework of the mean field approach, or global load sharing,
which means that after each fiber breaking the stress is equally
distributed on the intact fibers implying an infinite range of
interaction and a  neglect of stress enhancement in the
vicinity of failed regions
\cite{daniels,coleman,krajcinovic,sornette1,sornette2,bernardes,moukarzel,roux,yamir1,kun1,yamir2,phoenix1,phoenix2,phoenix3,phoenix4,curtin4,curtin5,curtin6}.
In spite of their simplicity FBM's capture the most important aspects of
material damage and they provide a deep insight 
into the fracture process.  
Over the past years several extensions of FBM have been carried out by
considering stress localization (local load transfer)
\cite{moukarzel,phoenix1,phoenix2,beyerlein,newman1,newman2,duxbury},  
the effect of matrix material between fibers
\cite{phoenix1,phoenix2,phoenix3,phoenix4,curtin4,curtin5,curtin6},
possible non-linear behavior of fibers \cite{krajcinovic},
thermally activated breakdown \cite{roux} of fibers, and coupling to an
elastic block \cite{delaplace}.

Very recently, a novel continuous damage law has been introduced 
in lattice models \cite{zvs} of fracture. In
this model when the failure threshold of a lattice bond is exceeded the
elastic modulus of the bond is reduced by a factor $a$ ($0 < a<1$), 
furthermore, multiple failures of bonds are allowed. Simulations
revealed that under strain controlled loading the system develops into
a self organized state, which is macroscopically plastic, and is
characterized by a power law distribution of avalanches of
breaks. We worked out an extensions of fiber
bundle models by implementing a continuous damage law for the fibers 
\cite{kun1}, in the spirit of Ref.\ \cite{zvs}. It has been
demonstrated in Ref.\ \cite{kun1} that the continuous damage fiber
bundle model (CDFBM)
provides a broad spectrum of description of materials varying its
parameters and for certain parameter settings the model recovers 
the variants of FBM known in the literature. 
CDFBM can be relevant for materials where 
the microscopic damage mechanism is gradual multiple failure of
components, {\it i.e.} matrix and fibers \cite{review,psh}. Very
recently, the CDFBM  
was further developed by Moral {\it et al.} taking into account time
dependence in the failure process \cite{yamir2}. 

One of the most appealing results  on CDFBM was 
that the multiple failure of brittle elements can give rise to a
macroscopic plastic behaviour of the specimen, which is then followed
by a hardening or softening regime, furthermore, under certain
conditions damage localization occurs.  However, the microscopic
damage process of CDFBM has not been explored.
The main goal of the present paper is to reveal the microscopic
failure process in order to understand the emergence of the
plastic macroscopic state. Analytic results are obtained to
characterize the damage process along the plateau under strain
controlled loading, furthermore, for stress controlled experiments we
develop a simulation technique and explore numerically the
distribution of bursts of fiber breaks. The effect of localization on
the process of damage is clarified. A phase diagram of the model
characterizing the possible constitutive behaviours and burst
distributions is constructed in terms of the two parameters of the model.

\section{model}

\begin{figure}
\begin{center}
\psfrag{aa}{\large $d_i^3$}
\psfrag{bb}{\large $d_i^1$}
\psfrag{cc}{\large $d_i^2$}
\psfrag{dd}{\large $d_i$}
\epsfig{bbllx=0,bblly=-15,bburx=400,bbury=500,file=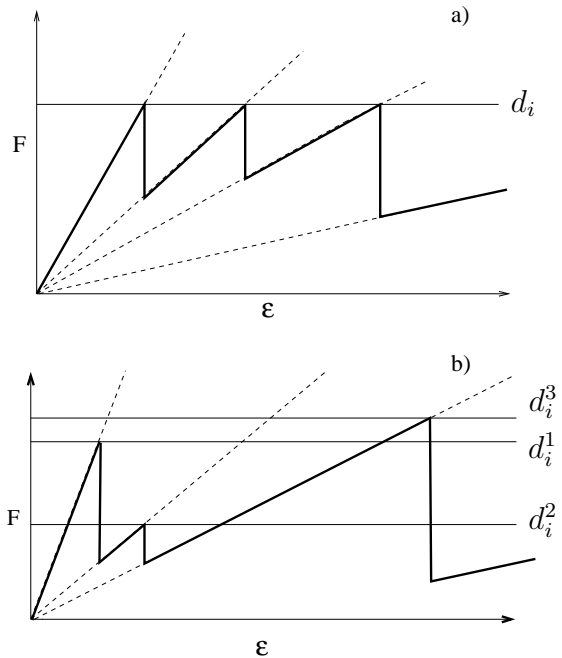,
  width=7cm}
\caption{The damage law of a single fiber of the
  continuous damage model when multiple failure is allowed $a)$ for
  quenched, and $b)$ for annealed disorder. The
  horizontal lines indicate the damage threshold $d_i$.}
\label{fig:damlaw}\end{center}
\end{figure}

The continuous damage fiber bundle model is an extension of the
commonly used fiber bundle models by
generalizing the damage law of fibers. 
The model system is composed of $N$ parallel fibers  with identical
Young-modulus $E_f$ but with 
random failure thresholds $d_i$, $i= 1, \ldots , N$. The failure strength
$d_i$ of individual fibers is an independent identically
distributed random variable with a probability density $p(d)$ and
a cumulative probability distribution $ P(d) = \int_0^d p(x) dx$.
The fibers are assumed to have linear elastic behaviour up to
breaking (brittle failure).
Under uniaxial loading of the specimen a fiber fails if it experiences
a load larger than its breaking threshold $d_i$. In the framework of
our model at the failure point the stiffness of the fiber gets reduced
by a factor $a$, where $0 \leq a < 1$, {\it i.e.} the stiffness of the
fiber after failure is $aE_f$. 
In principle, a fiber can now fail more than once and 
the maximum number $k_{max}$ of failures allowed for fibers is a
parameter of the model. Once a fiber has failed its damage threshold
$d_i$ can either be kept constant for the further breakings (quenched
disorder) or new failure thresholds of the same distribution can be
chosen (annealed disorder), which can model some microscopic
rearrangement of the material after failure. The damage law of the
model is illustrated in Fig.\ \ref{fig:damlaw} for both types of
disorder. 
The characterization of damage by  a continuous parameter corresponds to 
describe the system on length scales larger than the typical crack size.
This can be interpreted such that the smallest elements of the model 
are fibers and the continuous damage is due to cracking inside fibers.
However, the model can also be considered as the discretization of 
the system on length scales larger than the size of single fibers, so that
one element of the model consists of a collection of fibers with matrix 
material in between. In this case the microscopic damage mechanism 
resulting in multiple failure of the elements is the gradual cracking of 
matrix and the breaking of fibers.
In the following we refer to the elements of the continuous damage FBM
as fibers, but we have the above two possible
interpretations in mind. 

After failure the fiber skips a
certain amount of load which has to be taken by the other fibers. For
the load redistribution we assume infinite range of interaction among
fibers (mean field approach), furthermore, equal strain condition is
imposed which implies that stiffer fibers of the system carry more
load.  At a strain $\varepsilon$ the load of fiber $i$ that has failed
$k(i)$ times reads as
\begin{eqnarray}
\label{eq:fi}
f_i(\varepsilon) = E_fa^{k(i)} \varepsilon,
\end{eqnarray}
where  $E_fa^{k(i)}$ is the actual stiffness of fiber $i$. It is
important to note that, in spite of the infinite interaction range, Eq.\
(\ref{eq:fi}) is different from the usual 
global load sharing where all the intact fibers carry always the same
amount of load. In the following the initial fiber stiffness $E_f$ will be
set to unity.

\section{constitutive laws}
\label{sec:constit}

Here we provide a derivation of the constitutive law of the
continuous damage FBM in a more transparent way than 
in Ref.\ \cite{kun1}. This general
theoretical framework facilitates to obtain analytic results also for the
microscopic failure process. The key quantity is the probability
$pb_k(\epsilon)$ that during the loading of a specimen an arbitrarily
chosen fiber failed precisely $k$-times at a strain
$\epsilon$, where $k=0,\ldots, k_{max}$ denotes the failure index, and
$k=0$ is assigned to the intact fibers. $pb_k(\epsilon)$ can be cast
in the following form for {\it annealed disorder}
\begin{eqnarray}
\label{eq:pbi}
 && pb_k(\epsilon) = \left[1-P(a^k\epsilon)\right] \prod_{j=0}^{k-1}
P(a^j\epsilon), \\
\mbox{for} \quad && 0\leq k \leq k_{max}-1, \nonumber \\
\mbox{and} \quad && pb_{k_{max}}(\epsilon) = \prod_{j=0}^{k_{max}-1} 
P(a^j\epsilon), \nonumber
\end{eqnarray}
and for {\it quenched disorder}
\begin{eqnarray}
  \label{eq:pbq}
 && pb_0(\varepsilon) = 1-P(\varepsilon), \nonumber \\ 
 && pb_k(\varepsilon) = P(a^{k-1}\varepsilon) -  P(a^{k}\varepsilon), 
\ \mbox{for} \ 1\leq k \leq k_{max}-1,  \\ 
  && \mbox{and} \quad pb_{k_{max}}(\varepsilon)  =
P(a^{k_{max}-1}\varepsilon). \nonumber  
\end{eqnarray}

It can be easily seen that the probabilities Eqs.\
(\ref{eq:pbi},\ref{eq:pbq}) fulfill the normalization condition  
$\sum_{k=0}^{k_{max}} pb_k(\varepsilon) = 1$.
Average quantities of the fiber ensemble during a loading process can
be calculated using the 
probabilities Eqs.\ (\ref{eq:pbi},\ref{eq:pbq}). For instance, the
average load on a fiber $F/N$ at a given strain $\varepsilon$ 
reads as 
\begin{eqnarray}
  \label{eq:totload}
  \frac{F}{N} = \varepsilon \left[ \sum_{k=0}^{k_{max}} a^k
    pb_k(\varepsilon) \right],
\end{eqnarray}
which provides the macroscopic constitutive
behaviour of the model, and the expression in the brackets can be
considered as the macroscopic effective Young modulus of the
sample ($E_f= 1$). The single terms in the sum give the load carried
by the subset of fibers of failure index $k$. 
The variants of fiber bundle models used widespread in the literature can be
recovered by special choices of the parameters $k_{max}$ and $a$ of
the model. A micromechanical model of composites
\cite{phoenix3,phoenix4,curtin1,curtin2} can be obtained with
the parameter values $k_{max} = 1$, $a \neq 0$ 
\begin{eqnarray}
  \label{eq:micromech}
  \frac{F}{N} = \varepsilon \left[1-P(\varepsilon)\right] +
  a\varepsilon P(\varepsilon), 
\end{eqnarray}
while setting $k_{max} = 1$, $a = 0$, {\it i.e.} skipping the second
term in Eq.\ (\ref{eq:micromech}) results in the classical dry bundle
model of Daniels \cite{daniels}. 

\begin{figure}[htb]
\begin{center}
\epsfig{bbllx=135,bblly=165,bburx=470,bbury=700,
file=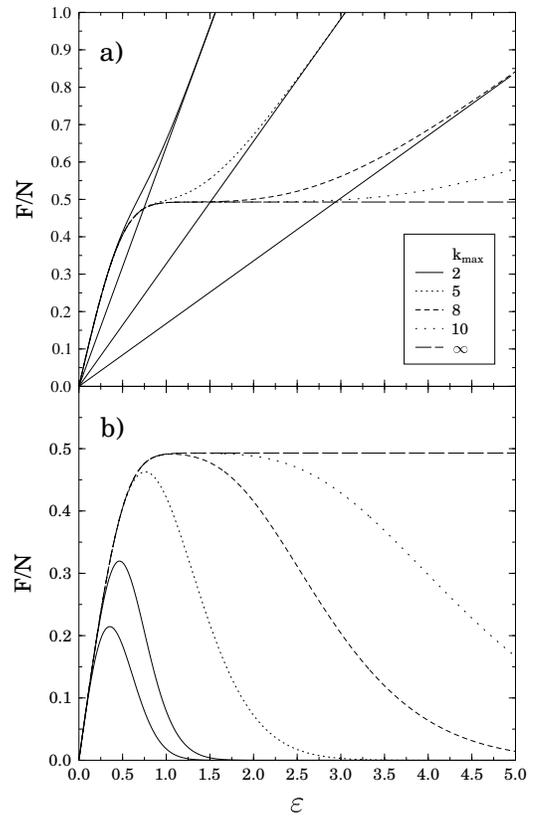,
  width=7cm}
\caption{Constitutive behaviour of the model of annealed disorder $a)$
  with $b)$ without 
  residual stiffness at $a=0.8$ for different values of $k_{max}$. In
  $b)$ the lowest curve presents the constitutive behaviour of the dry
  bundle model for comparison.}
\label{fig:constbehav}\end{center}
\end{figure}

In Fig.\ \ref{fig:constbehav}  we show the explicit form of the constitutive
law with annealed disorder for different values of $k_{max}$ in the
case of the Weibull distribution 
\begin{equation}
P(d)=1-\exp(-(d/d_c)^{m}),
\end{equation}
where $m$ is the Weibull modulus and $d_c$ denotes the characteristic
strength of fibers. The parameter values are set to $m=2$, $d_c=1$ in
all the calculations.

Note that in the constitutive equation Eq.\ (\ref{eq:totload}) the
term of the highest failure index $k_{max}$ can be conceived such
that the fibers have a residual stiffness of  $a^{k_{max}}$ after
having failed $k_{max}$ times. This residual stiffness results in a
hardening of the material, hence, 
the $F/N$ curves in Fig.\ \ref{fig:constbehav}(a) asymptotically 
tend to straight lines with a slope $a^{k_{max}}$. 
Increasing $k_{max}$ the hardening part of 
the constitutive behavior is preceded by a longer and longer 
plastic plateau, and in the limiting case of $k_{max}\to\infty$ the materials 
behavior becomes completely plastic. A similar plateau and asymptotic
linear hardening has been observed in brittle matrix composites, where
the multiple cracking of matrix turned to be responsible for the
relatively broad plateau of the constitutive behavior, and the
asymptotic linear part is due to the linear elastic behavior of
fibers remained intact after matrix cracking \cite{review}. 

In order to describe macroscopic cracking and global failure of a
specimen instead of hardening,
the residual stiffness of the fibers has to be set to zero 
after a maximum number $k^*$ of allowed failures \cite{kun1,zvs}. In
this case the constitutive law can be obtained from the general form 
Eq.\ (\ref{eq:totload}) by replacing $k_{max}$ in the upper limit of
the sum by $k^{*}-1$.
A comparison of the constitutive laws of the dry and continuous damage
FBM with global failure is presented  
in Fig.\ \ref{fig:constbehav}(b). One can observe that the dry 
FBM constitutive law has a relatively sharp maximum, while 
the continuous damage FBM curves exhibit a plateau whose length increases 
with increasing $k^*$. Note that  
the maximum value of $F/N$ corresponds to the macroscopic strength of the 
material, furthermore,  
in stress controlled experiments the plateau and the decreasing part  
of the curves cannot be reached. 
However, by controlling the strain $\varepsilon$, the plateau and the decreasing 
regime can also be realized. The value of the driving stress $\sigma
\equiv F/N$  corresponding to the plastic plateau, and the length of
the plateau are determined by
the damage parameter $a$, and  by $k_{max}$, $k^*$: Decreasing $a$ at
a fixed $k_{max}, k^*$, or increasing $k_{max}, k^*$ at a fixed $a$
give rise to an increase of the plateau's length.

\begin{figure}[htb]
\begin{center}
\epsfig{bbllx=145,bblly=405,bburx=480,bbury=700,
file=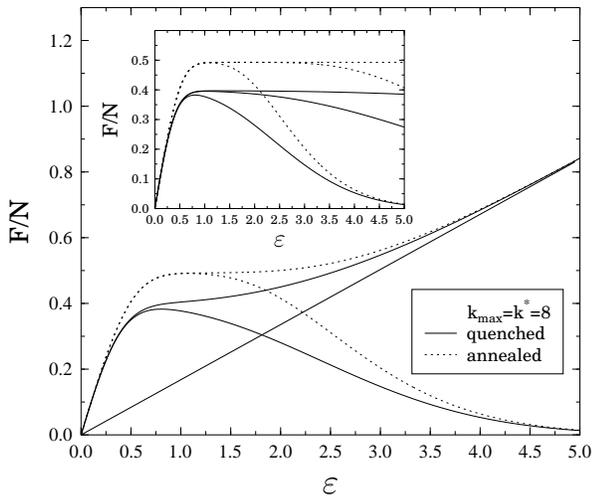,
  width=8cm}
\caption{Comparison of the constitutive behaviours with annealed and
  quenched disorder. We show data with and without remaining
  stiffness. The inset demonstrates how the shape of the constitutive
  curves changes when increasing $k^*$ with different types of disorders.
}
\label{fig:compare}\end{center}
\end{figure}

In Fig.\ \ref{fig:compare} we compare the constitutive laws of the model
with different types of disorder for hardening and softening of
fibers. It can be seen that there is no qualitative difference between
the curves of annealed and quenched disorder, however, the inset shows
that when the fibers do not have residual stiffness (softening) the
local shape of the curves around the maximum 
is different. The different types of maxima of the constitutive curve
of the two different disorders result in a very interesting behaviour
of the burst distributions, which will be discussed later. 

\section{Damage}

\begin{figure}[htb]
\begin{center}
\epsfig{bbllx=125,bblly=360,bburx=470,bbury=680,
file=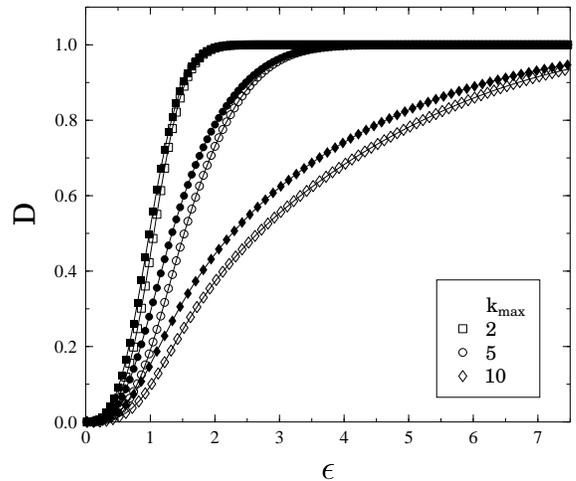,
  width=8cm}
\caption{The damage variable $D$ for annealed (open symbols) and
  quenched (filled symbols) disorder 
  for several different values of $k_{max}$. The damage variable was
  chosen to be $a=0.8$. 
}
\label{fig:damage}\end{center}
\end{figure}

The
damage state of the model at a certain $\varepsilon$ can be 
characterized by the average number of failures occurred. Based on the
probabilities Eqs.\ (\ref{eq:pbi},\ref{eq:pbq}), we introduce a
damage variable $D(\varepsilon)$ as
\begin{eqnarray}
  \label{eq:damage}
   D(\epsilon) = \frac{1}{k_{max}} \sum_{k=1}^{k_{max}} k pb_k(\epsilon),
\end{eqnarray}
which is an integral quantity of the damage process.
From the properties of $pb_k(\epsilon)$ it can be seen that $D$ is a 
monotonically increasing function, and $D\in[0,1]$.
Then the average number of failures can be obtained as $Nk_{max}
D(\epsilon)$.  
Fig.\ \ref{fig:damage} illustrates the behaviour of $D$ for three
different values of $k_{max}$.

\begin{figure}[htb]
\begin{center}
\epsfig{bbllx=145,bblly=340,bburx=470,bbury=640,
file=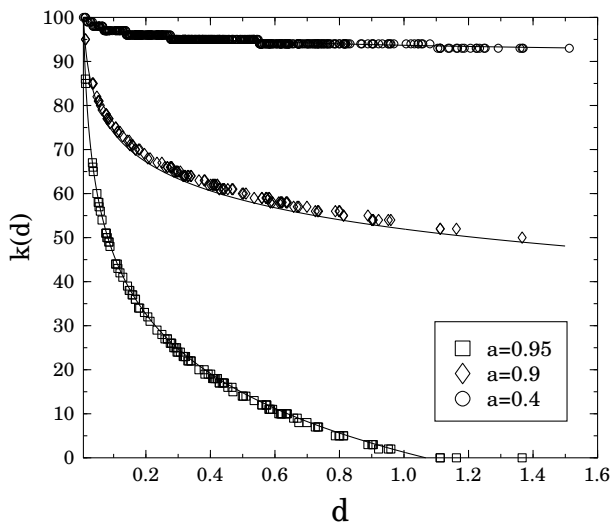,
  width=8cm}
\caption{The failure index $k$ of fibers as a function of their failure
  threshold $d$ for several different values of the damage parameter
  $a$. The number of fibers was chosen to be $N=500$. The continuous
  lines represent the corresponding analytical results of Eq.\
  (\ref{eq:failure}). 
}
\label{fig:localize}\end{center}
\end{figure}

It can be observed in Fig.\ \ref{fig:damage} that the overall
behaviour of the damage variable $D$ is nearly the same for 
annealed and quenched disorder, however, there is a significant
difference between the microscopic damage processes in the two
cases. In spite of the infinite range of interaction among fibers,
localization of damage occurs for the case of quenched disorder. It
means that weaker fibers tend to break more often than the stronger
ones. For quenched disorder, the strain $\varepsilon_m$  where the weakest 
fiber of failure threshold $d_m$ reaches $k_{max}$, is $\varepsilon_m
= d_m/a^{k_{max}}$. Hence, at this loading stage the  failure index
$k$ of fibers as a function of the damage threshold $d$ can be
obtained as  
\begin{eqnarray}
  \label{eq:failure}
k(d) = \frac{1}{\ln a} \ln \frac{d}{d_m}+k_{max}.
\end{eqnarray}
Localization of damage means that $k$ is a decreasing function of $d$,
and it can be seen from Eq.\  (\ref{eq:failure}) that the
localization gets more pronounced when the damage parameter $a \to 1$.
This localization effect is illustrated in Fig.\ \ref{fig:localize},
where the analytic result Eq.\ (\ref{eq:failure}) is compared to
simulations for three different values of $a$. 

\section{Distribution of bursts}
\label{sec:bursts}

One of the most interesting aspects of the damage mechanism of
disordered solids is that the breakdown 
is preceded by an intensive precursor activity in the form of avalanches
of microscopic breaking events \cite{hansen1,hansen2,hansen3,zrsv,zvs}. 
Under a given external load $F$ a certain fraction of fibers 
fails immediately. Due to the load transfer from broken to intact fibers
this primary fiber breaking may initiate secondary breaking that may
also trigger a whole avalanche of breakings. If $F$ is large enough the
avalanche does not stop and the material fails catastrophically.
For the dry FBM it has been shown by analytic means that in
the case of global load transfer the size distribution of
avalanches follows asymptotically a universal power law with an
exponent $-5/2$ \cite{hansen1,hansen2,zrsv}, however, in the case of local
load transfer no universal behaviour exists, and the avalanche
characteristic size is bounded \cite{hansen3}.  
This precursory activity can also be 
observed experimentally by means of the acoustic emission analysis.
Acoustic emission measurements have revealed that for a broad
variety of disordered materials the
response to an increasing external load takes place in
bursts having power law size distribution over a wide range
\cite{ciliberto,ae,zrsv}.

Introducing a continuous damage law in lattice models, simulations
revealed that under strain controlled conditions the system tends to a
steady state, which is macroscopically plastic \cite{zvs}, similarly
to our case. Due to the long range interaction, the plastic steady
state is characterized by power law 
distributed avalanches of breaks and it has been argued that the
underlying damage mechanism displays self organized criticality. In
the following we study the distribution of bursts in our 
CDFBM under strain and stress controlled conditions.

\subsection{Strain controlled case}
\label{sec:straincont}
Under strain controlled conditions of fiber bundles there is no load
transfer from broken to intact fibers since the load carried by each
fiber is determined by the externally imposed strain and the local 
fiber stiffness according to Eq.\ (\ref{eq:fi}). This implies that the
number of fibers which 
break due to an infinitesimal increase of the external strain is
completely determined by the statistics of fiber strength, {\em i.e.}
by $p(d)$ and $P(d)$.
It has been discussed in Sec.\ \ref{sec:constit} that the plastic
plateau and the decreasing part of the constitutive law can only be
realized in strain controlled experiments. To reveal the nature of
ductility arising in our model it turns to be useful to study the
statistics of bursts occurring under strain controlled conditions.

\begin{figure}[htb]
\begin{center}
\epsfig{bbllx=137,bblly=140,bburx=470,bbury=680,
file=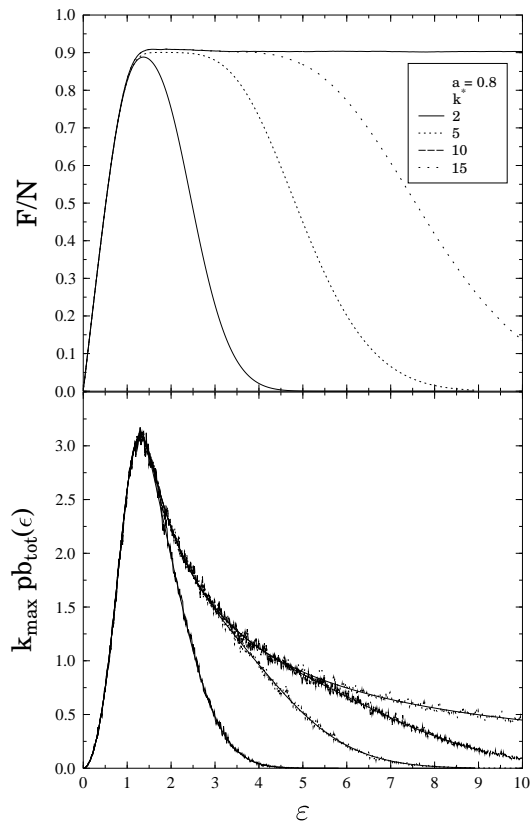,
  width=7cm}
\caption{$k_{max}pb_{tot}$ as a function of $\varepsilon$ for annealed
  disorder, comparison
  of simulations and analytic results of Eq.\ (\ref{eq:piia})
  (continuous lines).  The integral of the functions is always equal
  to $k^{*}$. In the upper part of figure the corresponding
  constitutive curves are also presented for comparison.
}
\label{fig:laveps}\end{center}
\end{figure} 
The basic quantity to characterize bursts is the probability
$pb_k^{k+1}(\epsilon) d\epsilon$ that a 
fiber, which has failed $k$ times up to strain $\epsilon$  imposed
externally, will fail again under an infinitesimal strain
increment  $d\epsilon$. From Eqs.\ (\ref{eq:pbi},\ref{eq:pbq})
$pb_k^{k+1}(\epsilon)$ can be cast in the form for {\it annealed disorder}
\begin{eqnarray}
  \label{eq:piia}
  pb_k^{k+1}(\varepsilon) &=& \prod_{j=0}^{k-1} P(a^j\varepsilon)
  \sum_{i=0}^{k} \frac{p(a^i \varepsilon)a^i}{P(a^i\varepsilon)}, \\
  k &=& 0, \ldots , k_{max}-1, \nonumber
\end{eqnarray}
and for {\it quenched disorder}
\begin{eqnarray}
  \label{eq:piiaq}
  pb_k^{k+1}(\varepsilon) = p(a^k\varepsilon)a^k,
 \qquad k=0, \ldots , k_{max}-1,
\end{eqnarray}
and the total probability of fiber breaking can be obtained by summing
over $k$
\begin{eqnarray}
pb_{tot}(\epsilon) &=& \frac{1}{k_{max}}\sum_{k=0}^{k_{max}-1}
pb_k^{k+1}(\epsilon).
\end{eqnarray}

\begin{figure}[htb]
\begin{center}
\epsfig{bbllx=137,bblly=140,bburx=470,bbury=680,
file=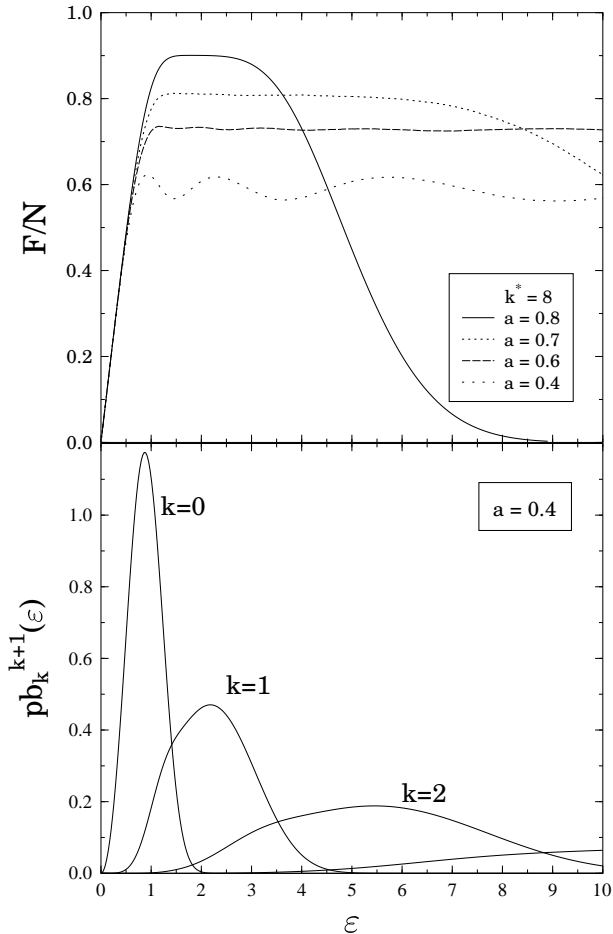,
  width=8cm}
\caption{$a)$ The constitutive behaviour varying the damage threshold
  at a fix $k^{*}$. $b)$ $pb_k^{k+1}(\varepsilon)$ for $a=0.4$.}
\label{fig:hasonaa}\end{center}
\end{figure} 
The number of fiber failures occurring in the strain interval
$[\epsilon, \epsilon + d\epsilon]$ can be obtained as
$N k_{max} pb_{tot}(\epsilon)d\epsilon$. This is a very important
characteristic quantity of the microscopic damage process since it can
be monitored experimentally by  means of acoustic emission
techniques. 
The behaviour of $pb_{tot}$ is shown in Fig.\ \ref{fig:laveps} for the
softening case with several values of $k^*$, where also the
corresponding constitutive curves are presented.  
It can be seen that $pb_{tot}$ has a
maximum where the plastic regime of the constitutive curve starts, and
it is a decreasing function of $\epsilon$ in the whole plastic
region. Due to the stiffness reduction of the system caused by the
subsequent failures, in the plastic regime the same increase of strain
results in smaller and smaller 
load increments on fibers, and hence, $pb_{tot}$ and the number of
failures decreases.  It also implies that the breaking activity, which
can be measured by acoustic emission techniques, decreases along the
plateau in agreement with experiments \cite{psh,pshtheory2,pshtheory3}. 

It follows from the above argument that decreasing the value of the
damage parameter 
$a$ while $k_{max}$ is kept fixed, the length of the plastic plateau,
preceding the decreasing or hardening part of the constitutive
behaviour, increases since larger strain is required to achieve
successive failure. This is demonstrated in Fig.\
\ref{fig:hasonaa}, where one can also see that for small $a$ the
constitutive curve develops distinct maxima. In order to clarify the
occurrence of these maxima in the plastic plateau, in Fig.\
\ref{fig:hasonaa} we also plotted $pb_k^{k+1}$ for three different
values of $k$ at $a=0.4$. With decreasing $a$ the length of the
plastic plateau increases, however, the consecutive maxima of
$pb_k^{k+1}$ get more and more separated giving rise to visible maxima
in the plateau. The broader the disorder distribution is, the smaller
the value of $a$ is where the maxima of $F/N$ appear.

\begin{figure}[htb]
\begin{center}
\epsfig{bbllx=125,bblly=360,bburx=470,bbury=680,
file=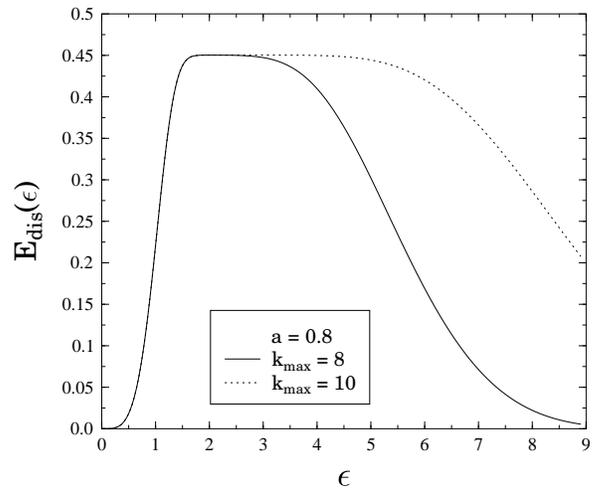,
  width=8cm}
\caption{The energy dissipation rate for two different values of
  $k_{max}$. }
\label{fig:energy}\end{center}
\end{figure} 
The energy dissipation rate is also a very important aspect of the
ductile regime of the model.
The energy dissipation rate $E_{dis}(\epsilon)$ is defined so that the
energy dissipated due to the 
failure of fibers in the strain interval  $[\epsilon, \epsilon + d\epsilon]$
can be obtained as $E_{dis}(\epsilon)d\epsilon$
\begin{eqnarray}
  \label{eq:edis}
  E_{dis}(\epsilon) = \sum_{k=0}^{k_{max}-1}
  \left[\frac{1}{2}\epsilon^2 a^k (1-a) \right]
  pb_k^{k+1}(\epsilon),
\end{eqnarray}
where the expression in the bracket provides the energy dissipated by
the failure of a fiber which has already failed $k$-times.
In Fig.\ \ref{fig:energy} the energy dissipation rate
$E_{dis}(\epsilon)$ is plotted for two different values of
$k_{max}$. Comparing Fig.\ \ref{fig:energy} to the corresponding
constitutive curves in Fig.\ \ref{fig:constbehav} it can be observed
that in the plastic regime $E_{dis}(\epsilon)$ is constant.

\subsection{Stress controlled case}
\label{sec:stresscont}

Under stress controlled loading conditions the microscopic dynamics of
the damage process is more complicated than in the strain controlled
case, since the failure of each fiber is followed by a redistribution
of load, which can provoke further fiber
breakings resulting in an avalanche of failure events. Studying the
statistics of avalanches under quasi-static loading of a specimen,
important information can be gained about the dynamics of damage,
which can be then compared to the results of acoustic emission
experiments. 
Due to the difficulties of the analytic treatment, we develop a
simulation technique and explore numerically the properties  
of bursts in our continuous damage fiber bundle model. The
interaction of fibers, the way of load redistribution
is crucial for the avalanche activity. A very important property 
of CDFBM is that in spite of the infinite range of interaction the
load on intact fibers is not equal, but stiffer fibers cary more
load, furthermore, for quenched disorder damage localization occurs,
which might affect also the avalanche activity. 

To implement the quasi-static loading of a specimen of $N$ fibers in
the framework of CDFBM, the local load on the fibers $f_i$ has to be
expressed in terms of the external driving $F$. 
Making use of Eq.\ (\ref{eq:fi}) it follows that
\begin{eqnarray}
  \label{eq:finew}
  F = \sum_{i=1}^{N} f_i = \epsilon  \sum_{i=1}^{N} a^{k(i)},
\end{eqnarray}

\noindent 
and hence, the strain and the local load
on fibers can be obtained as

\begin{eqnarray}
  \label{eq:stresscont}
  \epsilon = \frac{F}{\sum_{i=1}^{N} a^{k(i)}}, \quad f_i = F \frac{
      a^{k(i)}}{\sum_{i=1}^{N} a^{k(i)}},
\end{eqnarray}
when the external load $F$ is controlled.
The simulation of the quasi-static loading proceeds
as follows: in a given stable state of the system we determine the load
on the fibers $f_i$ from the external load $F$ using Eq.\
(\ref{eq:stresscont}). The next fiber to break can be found as
\begin{eqnarray}
  \label{eq:nextb}
  r = \min\limits_{{\displaystyle{i^{*}}}} \frac{d_i}{f_i}, \qquad r > 1,
\end{eqnarray}
{\it i.e.} that fiber breaks for which the ratio $d_i/f_i$ is the
smallest. Here $i^{*}$ denotes the index of the fiber to break, $d_i$ is
the damage threshold of fiber $i$, and $f_i$ is the local load on
it. To ensure that the local load of a fiber is proportional to
its stiffness, the external load has to be increased in a multiplicative
way, so that $F \rightarrow rF$ is imposed, and the failure
index of fiber $i^{*}$ is increased by one $k(i^{*}) \rightarrow 
k(i^{*})+1$. After the breaking of fiber $i^{*}$, the load $f_i$
carried by the fibers  
has to be recalculated making use of Eq.\ (\ref{eq:stresscont}), which 
provides also the correct load redistribution of the model. If there are
fibers in the state obtained, whose load exceeds the local breaking
threshold, they fail, {\it i.e.} their failure index is
increased by 1 and the local load is again recalculated until a stable
state is obtained.  A fiber cannot break any longer if its failure
index $k$ has reached $k^{*}$ or $k_{max}$ during the course of the
simulations. 
This dynamics gives rise to a complex avalanche activity of fiber
breaks, which is also affected by the type of disorder. The size of an
avalanche $S$ is defined as the number of breakings initiated by a single
failure due to an external load increment. 
 
Simulations revealed that varying the two parameters of the model
$k_{max}, a$, or $k^{*}, a$ and the type of disorder, 
the CDFBM shows an interesting variety of avalanche activities,
characterized by different shapes of the avalanche size
distributions. 
\begin{figure}
 \begin{center}
  \epsfig{bbllx=140,bblly=400,bburx=500,bbury=700,
  file=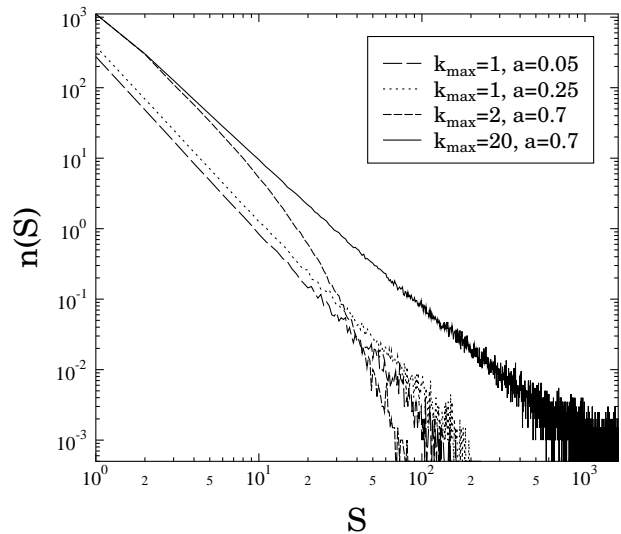,
  width=9cm}
  \caption{Avalanche size distributions for different values of
    $k_{max}$ and $a$ when fibers have remaining stiffness and the
    disorder is annealed. The number of fibers was $N=1600$ and
    averages were made over 2000 samples. The number of
    avalanches $n$ of size $S$ are  
    shown to demonstrate also how the total number of avalanches changes.} 
  \label{fig:aval_an_r}
 \end{center}
\end{figure}
In Fig.\ \ref{fig:aval_an_r} the histograms $n(S)$ of the 
avalanche sizes $S$ are shown which were obtained for a system of
remaining stiffness and annealed disorder with  Weibull parameters
$m=2$, $d_c=1$. Since in the limiting case 
of $a \rightarrow 0$ the CDFBM recovers the global load sharing dry fiber 
bundle model, in Fig.\ \ref{fig:aval_an_r} the curves with small $a$
and $k_{max}=1$ are power laws with an exponent $\alpha=5/2$ in
agreement with the analytic results \cite{hansen1,hansen2}. Increasing 
the value of $a$ at a fixed $k_{max}$ only gives rise to a larger number of
avalanches, {\it i.e.} parallel straight lines are obtained on a
double logarithmic 
plot, but the functional form of $n(S)$ does not change. However, when 
$a$ exceeds a critical value $a_c$ ($a_c \approx 0.3$ was obtained
with the Weibull parameters specified above) the avalanche
statistics drastically changes. At a fixed $a > a_c$ when
$k_{max}$ is smaller than a specific value $k_c(a)$,
the avalanche sizes show exponential distribution, while above
$k_c(a)$ the distribution takes a power law form with an
exponent $\beta = 2.12 \pm 0.05$. 

\begin{figure}
\begin{center}
\psfrag{aa}{\large $a$}
\psfrag{bb}{\large 1}
\psfrag{cc}{\large $a_c$}
\psfrag{ii}{\large $0$}
\psfrag{dd}{\large 1}
\psfrag{ee}{\large $k_{max}$}
\psfrag{ff}{\LARGE{ $\sim e^{-s/s_o}$}}
\psfrag{gg}{\LARGE $\sim  s^{-\beta}$}
\psfrag{hh}{\LARGE $\sim s^{-\alpha}$}
\psfrag{kk}{\LARGE {DBM}}
\epsfig{bbllx=0,bblly=-15,bburx=315,bbury=235,file=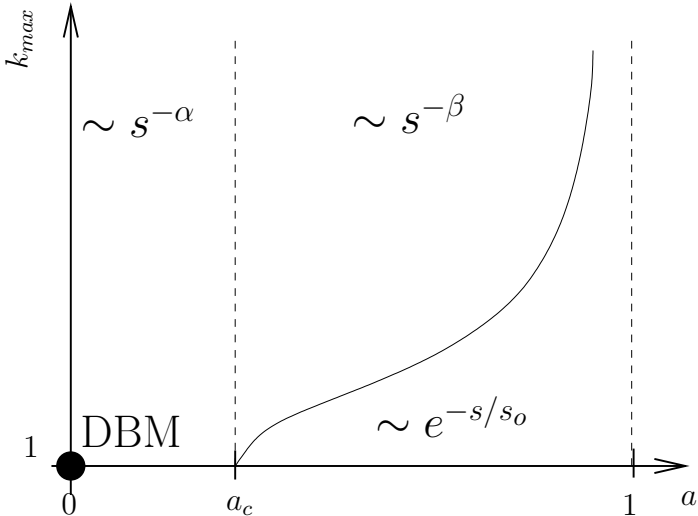,
  width=9cm}
\caption{Phase diagram for the continuous damage model with remaining
  stiffness for both types of disorder. The functional form of the
  avalanche statistics  is given in the parameter regimes. The location
  of the Dry Bundle Model (DBM) in the parameter space is also indicated.} 
\label{fig:phases}
\end{center}
\end{figure}
Based on the above results of
simulations a phase diagram is constructed which summarizes the
properties of avalanches with respect to the parameters of the model.
Fig.\ \ref{fig:phases} demonstrates the existence of 
three different regimes. If the damage  parameter $a$ is smaller than 
$a_c$,  the dynamics of avalanches is
close to the simple Dry Bundle Model characterized by a power law of
the mean field exponent $\alpha = 5/2$. However, for $a>a_c$ the
avalanche size distribution depends on the number of
failures $k_{max}$ alowed. 
The curve of $k_c(a)$ in the phase diagram
separates two different  regimes. For the parameter regime below
the curve, avalanche  distributions with an exponential shape were obtained.  
However, the parameter regime above  $k_c(a)$ is characterized by a
power law distribution of avalanches with a constant exponent $\beta
= 2.12 \pm 0.05$ significantly different from the mean field exponent
$\alpha=5/2$ 
\cite{hansen1,hansen2}. It is important to emphasize that the overall
shape of phase diagram is independent of the type of the  disorder
(annealed or quenched), moreover, the specific values $a_c \approx
0.3$ and $k_c(a)$ depend on the details of the disorder distribution
$p(d)$.  

A very different behavior was obtained for the system when fibers do
not have remaining stiffnes after a $k^{*}$ number of failures.  
Simulations revealed that in this case the
avalanche statistics strongly depends on the type of disorder. 
When the disorder is quenched the size distribution of avalanches
follows always the dry bundle results
for the whole domain of parameters, {\it i.e.} $N(S)$ shows power law
behaviour with an exponent $\alpha = 5/2$. When $k^{*} > 1$ the
larger number of breakings results in more avalanches but the overall
distribution does not change. 
\begin{figure}
  \begin{center}
    \epsfig{bbllx=145,bblly=400,bburx=500,bbury=700,
     file=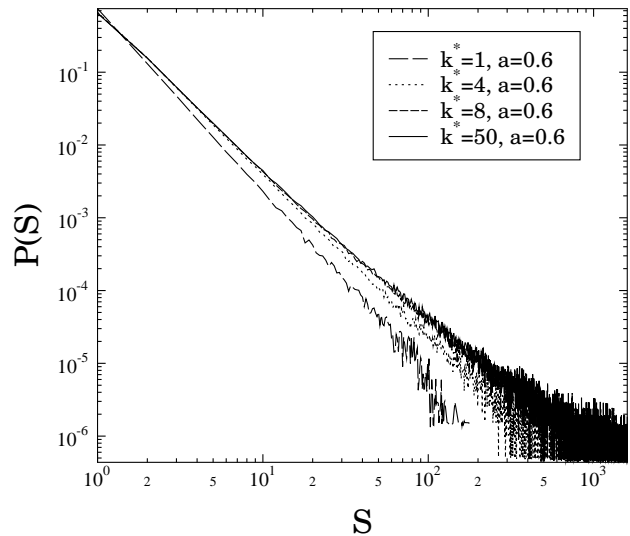,
     width=9cm}
     \caption{Avalanche size distributions for different values of
       $k^*$ at a fixed $a > a_c$ when fibers have no remaining
       stiffness and the disorder is annelaed. 
       }
     \label{fig:aval_an}
  \end{center}
\end{figure}
Nevertheless, when the disorder is annealed the system 
shows a more  complex behaviour.
When  $a$ falls below a certain critical value $a_c$ the results are
similar to DBM independently  of the value of
$k^{*}$, however, for $a>a_c$ a novel avalanche dynamics
appears (for the present values of the Weibull parameters $a_c \approx 
0.35$ was obtained).         
In Fig.\ \ref{fig:aval_an} the avalanche distributions are shown for
an $a$ value above $a_c$, varying the value of $k^{*}$. 
It is very important to emphasize that the curves in all the cases can 
be well fitted with a power law, however, the value of the exponent
depends on  $k^{*}$.
Two extreme cases can be distingushed: for $k^*=1$  the
system recovers the DBM avalanche dynamics.
On the other hand, for $k^*>k_c(a)$ the exponent of the power laws
is $\beta = 2.12\pm 0.05$, similarly to the case
of remaining stiffness. Below $k_c(a)$ the exponents vary as a
function of $k^*$ between the
mean field exponent $\alpha$ and $\beta$. 

Based on Refs.\ \cite{hansen1,hansen2} the different types of
avalanche size distributions can also be understood up to some extent in
terms of the constitutive curves of Sec.\ \ref{sec:constit}. Comparing 
Fig.\ \ref{fig:compare} and Figs.\ \ref{fig:aval_an_r}, \ref{fig:aval_an} it
can be recognized that if the constitutive curve has a single
quadratic maximum the corresponding avalanche size distribution of
CDFBM follows the mean field results, while other types of avalanche
statistics arises when this condition does not hold.  

\section{Conclusions}
A detailed analytical and numerical study of the continuous damage
fiber bundle model is presented. The model is an extension of the
classical fiber bundle model by introducing a continuous damage law,
and allowing for multiple failure of fibers with quenched and annealed 
disorders. A simple general derivation of the constitutive behaviour
of the model is provided, which also facilitates to obtain analytic
results for the microscopic damage process. Varying its parameters,
the model 
provides a broad spectrum of description of materials ranging from
strain hardening to perfect plasticity, and hence, the model can be
relevant to describe the damage process of various types of materials
\cite{review,pshtheory1,psh,pshtheory2,pshtheory3}.  
It is a remarkable feature of the model that multiple failure of
brittle elements can result in a macroscopically plastic state, which
has also been observed experimentally in materials where the damage
mechanism is the gradual multiple failure of ingredients
\cite{pshtheory2,pshtheory3}. 
 
 The present study focused on the microscopic damage process to
understand the emergence of the plastic plateau under strain
controlled loading, and the resulted avalanche activity under stress
controlled loading of the continuous damage fiber bundle
model. Analytic results are obtained to
characterize the damage process along the plateau under strain
controlled loading, furthermore, for stress controlled experiments 
a simulation technique was developed and the distribution of
avalanches of fiber breaks was explored numerically. Simulations
showed that depending on the parameters of the model the distribution
of bursts of fiber breaks can be exponential or power law.
Based on extensive computer simulations, a phase diagram
characterizing the possible avalanche distributions is constructed in
terms of the two parameters of the model. One of the most appealing
outcomes is that the model has a broad parameter regime where the
avalanche statistics shows a power law behaviour with an exponent
significantly different from the well know mean field exponent, in
spite of the infinite range of interaction among fibers. 
The results obtained have relevance to understand the acoustic
emission measurements performed on various elasto-plastic materials
\cite{review,pshtheory1,psh,pshtheory2,pshtheory3}.

\section*{Acknowledgment}
This work was supported by the project SFB381, and by the NATO grant
PST.CLG.977311.    
F.\ Kun  acknowledges financial support of the
B\'olyai J\'anos Fellowship of the Hungarian Academy of Sciences and
of the Research Contract FKFP 0118/2001.

\end{multicols}
\end{document}